\documentclass[5p,twocolumn,times,number]{elsarticle}

\usepackage{graphicx}
\usepackage{amsmath}   
\usepackage{microtype}   
\usepackage{lineno}

\begin{document}

\begin{frontmatter}

\title{Radio detection of high-energy cosmic rays with the Auger Engineering Radio Array}

\author[add1]{Frank~G.~Schr\"oder\corref{cor}}
\ead{frank.schroeder@kit.edu}
\author[add2]{for the Pierre Auger Collaboration}

\cortext[cor]{Corresponding author}

\address[add1]{Institut f\"ur Kernphysik, Karlsruhe Institute of Technology (KIT), Karlsruhe, Germany}
\address[add2]{Observatorio Pierre Auger, Malarg\"ue, Argentina, (full author list: http://www.auger.org/archive/authors\_2015\_06.html)}

\begin{abstract}
The Auger Engineering Radio Array (AERA) is an enhancement of the Pierre Auger Observatory in Argentina. Covering about $17\,$km\textsuperscript{2}, AERA is the world-largest antenna array for cosmic-ray observation. It consists of more than 150 antenna stations detecting the radio signal emitted by air showers, i.e., cascades of secondary particles caused by primary cosmic rays hitting the atmosphere. At the beginning, technical goals had been in focus: first of all, the successful demonstration that a large-scale antenna array consisting of autonomous stations is feasible. 
Moreover, techniques for calibration of the antennas and time calibration of the array have been developed, as well as special software for the data analysis. Meanwhile physics goals come into focus. At the Pierre Auger Observatory air showers are simultaneously detected by several detector systems, in particular water-Cherenkov detectors at the surface, underground muon detectors, and fluorescence telescopes, which enables cross-calibration of different detection techniques. 
For the direction and energy of air showers, the precision achieved by AERA is already competitive; for the type of primary particle, several methods are tested and optimized. By combining AERA with the particle detectors we aim for a better understanding of cosmic rays in the energy range from approximately $0.3$ to $10\,$EeV, i.e., significantly higher energies than preceding radio arrays.    
\end{abstract}


\end{frontmatter}

\section{Introduction}
The Auger Engineering Radio Array (AERA) detects cosmic-ray air showers by their radio emission. It is one of the enhancements of the Pierre Auger Observatory in Argentina \cite{AugerDespricption2015}, and recently was extended in size. It now consists of 153 antenna stations spread with different spacings over an area of approximately $17\,$km\textsuperscript{2}. Thus, it currently is the largest antenna array dedicated to cosmic-ray measurements.

The Pierre Auger Observatory consists of several detector systems for air showers covering an area of $3000\,$km\textsuperscript{2}, aiming at studying the highest-energy cosmic rays. The two main detectors are an array of 1660 surface detectors measuring secondary air-shower particles, and 27 telescopes overlooking the array and observing fluorescence emission caused by the air showers in the atmosphere. The surface detectors are water-Cherenkov detectors, and will soon be upgraded adding a scintillation detector on top \cite{AugerUpgrade_PISA2015}. 
By this upgrade the accuracy for the separation of secondary muons and electrons of the air shower will be increased. This is important, since the size ratio of the electromagnetic and muonic air-shower components provides one of the two most important estimators of the type of the primary cosmic-ray particles. Another estimator is the atmospheric depth of the shower maximum, which is obtained from the fluorescence measurements, but only during dark nights with clear sky.

\begin{figure*}[t]
  \centering
  \includegraphics[width=0.8\linewidth]{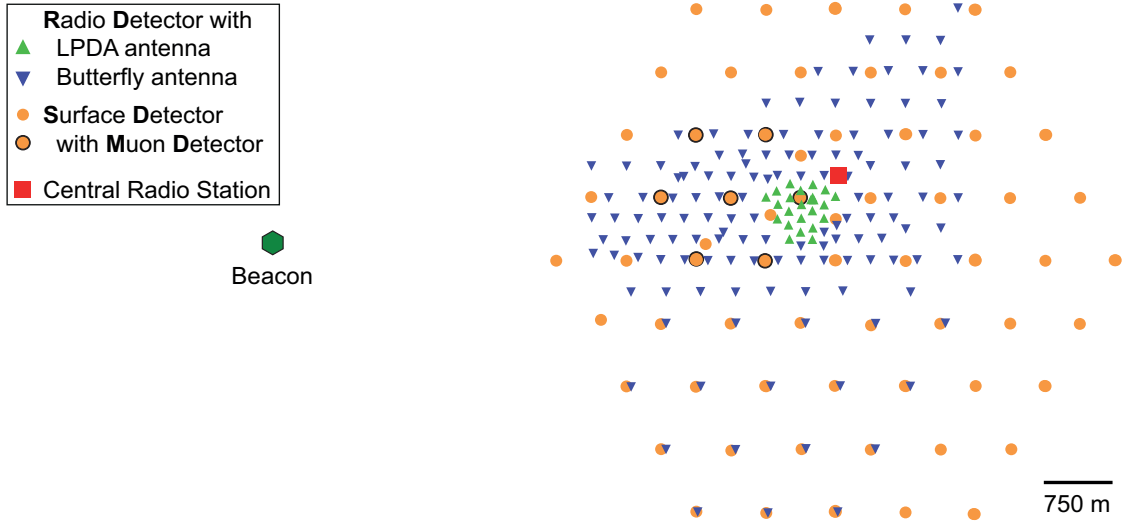}
  \caption{Map of the enhancement area of the Pierre Auger Observatory consisting mainly of the Auger Muon and Infill Ground Array (AMIGA) and the Auger Engineering Radio Array (AERA). Each side of the hexagon is about $3\,$km long, and the spacing between the surface detectors for the air-shower particles is $750\,$m. AERA features radio detectors with antennas of different types, a reference beacon, and a central radio station for data-acquisition and monitoring purposes.}
  \label{fig_map}
\end{figure*}

In an enhancement area, the Pierre Auger Observatory features additional detectors (see Fig.~\ref{fig_map}): the Auger Muon and Infill Ground Array (AMIGA) has a reduced spacing of the surface detectors to lower its energy threshold close to $10^{17}\,$eV. For some of the surface detectors additional underground scintillation counters are installed. These counters enable a more accurate measurement of the muon number, which can be exploited for air-shower analyses as well as for calibration of the upgraded surface detectors. AERA covers a large part of AMIGA, and in particular the full area already equipped with muon counters.

This setup enables simultaneous measurements of the same air showers with different detectors yielding complementary information. Like the fluorescence signal, the radio signal is generated by the electromagnetic component of the air shower during the shower development, yet, by a different mechanism. The radio emission mainly originates from the geomagnetic deflection of the electrons and positrons \cite{KahnLerche1966}, which is coherently amplified at wavelengths of a few meters, since this is the typical thickness of the shower front. Askaryan emission \cite{Askaryan1962}, i.e., radio emission due to the time-varying net-charge excess of the shower also contributes, but is almost an order of magnitude weaker. The surface detectors are sensitive to the muonic and electromagnetic shower components, the muon counters only to the muonic component. Hence the combination of the different detectors should enhance the total accuracy of the air-shower measurements, which is investigated experimentally comparing 
AERA and 
AMIGA with fluorescence measurements.

AERA is a multi-purpose experiment aiming at several technological and scientific goals:
\begin{itemize}
 \item \textbf{Demonstrating the technical feasibility of a large-scale radio array consisting of remote antenna stations.} For this purpose, AERA features different type of antennas, station electronics, data-acquisition systems, and layouts for the antenna array. Thus, the advantages and shortcomings of the different configurations can be compared. For large scale applicability, a radio array has to balance physics features, e.g., aiming at maximum accuracy and lowest possible threshold, with economic and manpower resources. Hence, AERA enables the community to get a realistic estimate of the demands and possibilities.

 \item \textbf{Calibration of the radio signal emitted by air showers.} Knowing all features of the radio signal is the basis for the reconstruction of the air-shower properties, which in particular are the direction, the energy, and the position of the shower maximum. The latter is a key observable to statistically estimate the mass composition of the primary cosmic particles. The understanding of the signal properties has to be sufficiently deep, to compete in the precision for the air-shower parameters with other detection techniques. Probably, this goal has already been achieved, although with respect to the shower maximum, the experimental proof still has to be delivered, e.g., by comparing the radio values for the shower maximum to the measurements of the fluorescence telescopes.
 
 \item \textbf{Cosmic-ray science in the energy range from a few $10^{17}\,$eV to $10^{19}\,$eV.} This is approximately the range from the second knee to the ankle in the cosmic-ray energy spectrum, where a transition from galactic to yet unknown extra-galactic cosmic rays is expected. AERA will provide measurements of the energy and estimations for the mass composition in this energy range. Hybrid analyses combining the AERA measurements with the particle and fluorescence measurements have the potential to enhance the total accuracy compared to the individual performance of the detectors. This will help to better discriminate between different theoretical models for the transition from galactic to extragalactic cosmic rays.

\end{itemize}

\section{Detector Description}
\begin{figure}[t!]
  \centering
  \includegraphics[width=0.99\linewidth]{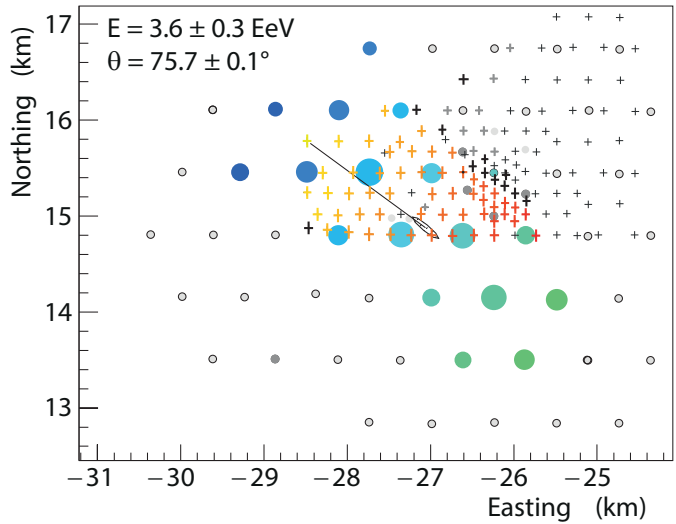}
  \caption{Footprint of an inclined event. Colored circles are particle detectors with signal, crosses are antenna stations, where in both cases the color code indicates the arrival time of the signal. The shower direction and core is indicated by the line with the ellipse.}
  \label{fig_footprint}
\end{figure}

\begin{figure}
  \centering
  \includegraphics[width=0.99\linewidth]{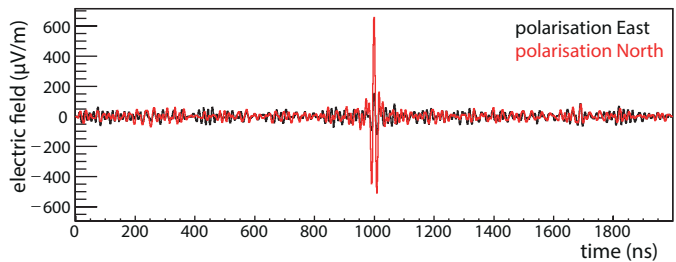}
  \caption{Time series of the east-west and north-south polarization of the electric field recorded at one station of the example event shown in figure \ref{fig_footprint}. The radio pulse has a typical width of a few $10\,$ns, where the pulse shape is mainly determined by the response of the analog bandpass filter in the signal chain.}
  \label{fig_trace}
\end{figure}

Before AERA, the feasibility of the digital radio-detection technique had already been proven by smaller experiments at lower energies, namely LOPES \cite{FalckeNature2005} and CODALEMA \cite{ArdouinBelletoileCharrier2005}. Therefore, it is a technological success by itself, that AERA by now detected more than $5000$ air showers using a sparse antenna spacing of $150\,$m and more (see Figs.~\ref{fig_footprint} and \ref{fig_trace} for an example event). For this, AERA uses a frequency range similar to other experiments, namely $30-80\,$MHz, since this band provides a good signal-to-noise ratio at distances of a few $100\,$m from the shower axis. Each AERA station features at least two orthogonal antennas, which enables a reconstruction of the electric-field vector.

In contrast to predecessors, AERA stations are autonomous and remote. This means that each station features its own local power supply via solar panel and battery, and its own local analog and digital electronics, which records and buffers the time series of the electric field measured by the antennas. Upon request, individual stations transfer the measured data to a central data-acquisition using wireless communication.

\begin{figure}[t!]
  \centering
  \includegraphics[width=0.99\linewidth]{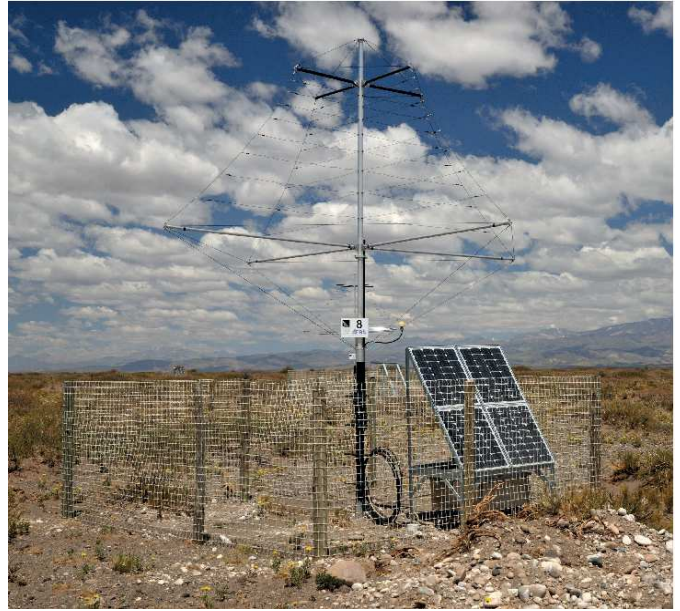}
  \caption{AERA station with logarithmic periodic dipole antenna (LPDA).}
  \label{fig_lpda}
\end{figure}

\begin{figure}
  \centering
  \includegraphics[width=0.99\linewidth]{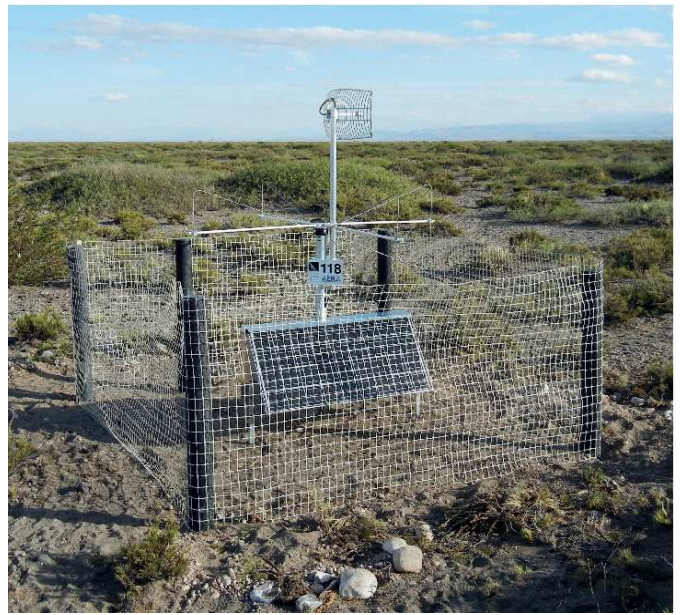}
  \caption{AERA station with Butterfly antenna.}
  \label{fig_butterfly}
\end{figure}

At AERA mainly two different types of antennas are used: logarithmic periodic dipole antennas (LPDAs) (Fig.~\ref{fig_lpda}), which feature a gain pattern relatively independent from ground conditions, and butterfly antennas (Fig.~\ref{fig_butterfly}), which are easier to deploy and feature a higher gain. Furthermore, different prototypes for the vertical component of the electric field are currently under test at AERA, which could further enhance the sensitivity for inclined showers.

The effect of the antennas on the amplitudes and phases of measured signals is taken into account during analysis. For this purpose we use dedicated simulations as well as calibration measurements of the antennas \cite{AERA_AntennaPaper2012}. Since also the properties of all analog components, i.e., filters, amplifiers and cables, are known, the measured electric field can be reconstructed in physical units (V/m) at each antenna station.

Moreover, a consistent amplitude and time calibration over the whole array is important. Slight differences in the gain of individual stations are studied using galactic noise regularly recorded as background. Nanosecond-level time synchronization is provided using a reference beacon as pioneered by LOPES \cite{SchroederTimeCalibration2010}. The beacon continuously emits sine waves, whose relative phasing in the individual stations is used for an offline improvement of the relative timing accuracy from a few $10\,$ns to better than $2\,$ns. This is crucial for reconstructing the shape of the radio wavefront \cite{LOPES_Wavefront2014, LOFAR_Wavefront2015}, which, on a statistical level, contains information on the mass composition of the primary cosmic rays.

For AERA, the analysis software framework of the Pierre Auger Observatory has been extended \cite{RadioOffline2011}. Since the software framework features a modular structure, it is easy to adapt for different analysis needs. Since the functionality is of general relevance for any digital radio array, it has been made available for the community, and is already in use by other experiments, e.g., Tunka-Rex \cite{TunkaRex_RICAP2013}.

\section{Results}
The first results of AERA concern the technical feasibility, and the properties of the radio signal. The general picture for the radio emission could be confirmed (cf.~Refs.~\cite{Werner2012, AlvarezMuiz2012, Ludwig2011, Marin2012} for recent theoretical results). For the first time, in addition to the east-west polarized geomagnetic emission, a radially polarized component could be identified in the measured events, which has an average relative strength of $14\,\%$ \cite{AERA_Polarization2014}. This is expected due to the Askaryan effect, which contributes to the radio emission in addition to the geomagnetic effect. Additional evidence for this had been found by LOFAR \cite{LOFARpolarization2014} in the polarization signal and by CODALEMA \cite{CODALEMA_LDFassymetry2015} in the amplitude signal, since the interference of the geomagnetic and Askaryan effects leads to an azimuthal asymmetry of the lateral distribution. Within the measurement uncertainties, AERA events seem to be consistent with the 
prediction of various simulation codes, which predict such an asymmetric lateral distribution.

The current focus of AERA analysis is on optimizing methods for the reconstruction of air-shower parameters, in particular the energy and the shower maximum. Since the radio technique measures the energy of the electromagnetic shower component in a calorimetric way, the achievable accuracy should be similar to the accuracy of the fluorescence technique. First results indicate that the achieved precision is at least similar to the precision achieved with the surface detectors \cite{AERAenergy2012}. Currently the reconstruction method is further optimized taking into account the asymmetry of the lateral distribution by fitting a two-dimensional lateral-distribution function to the detected radio footprint. A direct comparison to the fluorescence technique, however, requires larger statistics of corresponding hybrid events, which should be available soon.

The two main shower properties sensitive to the mass of the primary cosmic-ray particles are the atmospheric depth of the shower maximum, $X_\mathrm{max}$, and the relative sizes of the electromagnetic and muonic shower components, the electron-muon ratio. Several properties of the radio signal can be used for the reconstruction of the shower maximum, and the corresponding methods are currently under development and test at AERA.

Two methods rely on the shape of the lateral distribution, either fitting the lateral distribution with a predefined function containing parameters sensitive to $X_\mathrm{max}$ \cite{LOPES_PRD2012}, or comparing several shower simulations with different $X_\mathrm{max}$ values to the measured event \cite{LOFAR_Xmax2014}. In general, the closer the shower maximum to the detector, the steeper is the lateral distribution. The same general argument is true for the shape of the radio wavefront of the shower, which provides an additional method using the measured arrival times in the individual antennas \cite{LOPES_Wavefront2014}. Finally, the shape of the frequency spectrum is sensitive to the shower maximum \cite{AERAspectrum2012}, which potentially would provide a way to get information on $X_\mathrm{max}$ in a single antenna, while the previous methods require a detection in at least three or four antenna stations.

Radio detection is also important for the second mass-sensitive parameter, since the radio signal provides a measurement of the size of the electromagnetic component, and the surface detectors also measure the muonic shower component. Thus, a combination of the radio and particle measurements would provide complementary information on the mass composition. This method is especially interesting for inclined showers. First, for inclined showers the radio footprint becomes large \cite{SaftoiuLOPESinclined2009}, i.e., of comparable size as the particle footprint, making sparse radio arrays economic. Second, for inclined showers the electromagnetic component dies out in the air, and only muons arrive on ground. Consequently, combining particle and radio detectors seems to be the only way to measure the electron-muon ratio for inclined showers.

\section{Conclusion and Outlook}
AERA is the largest radio array for air-shower detection, and demonstrated that successful detection is possible with an antenna spacing of several $100\,$m. Its latest extension has an even coarser spacing of $750\,$m, which is equivalent to the spacing of the particle detectors, and should be sufficient for the detection of inclined showers. First results of AERA show that the understanding of the radio emission has dramatically improved in the last years, meaning that within all measurement uncertainties no significant discrepancies between AERA measurements and theoretical descriptions have been found so far, and more thorough tests are required.

For the reconstruction of the shower direction and energy first results are promising and indicate that AERA can achieve a precision of at least the same level as the surface detectors. The next years will show whether the accuracy for both the energy and the shower maximum can compete even with the fluorescence detectors. Moreover, we currently investigate the potential which lies in the complementarity of the simultaneous radio and particle measurements. Enhancing the overall accuracy for the shower properties will help to better understand the transition from galactic to extragalactic cosmic rays, which is expected in the energy range of AERA.


\end{document}